\pdfoutput=1

\documentclass[11pt]{article}
\usepackage{EMNLP2022}
\usepackage{hyperref}
\usepackage{times}
\usepackage{latexsym}
\usepackage{graphicx}
\usepackage{subfigure}
\usepackage{multirow}
\usepackage{booktabs}
\usepackage{algorithm}
\usepackage{algpseudocode}
\usepackage{amsmath}
\usepackage{graphbox}
\usepackage{multirow}
\usepackage{amssymb}
\usepackage{arydshln}
\usepackage{bbm}
\renewcommand{\algorithmicrequire}{\textbf{Input:}}  
\usepackage{float}
\usepackage[T1]{fontenc}

\usepackage[utf8]{inputenc}

\usepackage{microtype}

\usepackage{inconsolata}

\usepackage{xcolor}

\usepackage{caption}
\newcommand{\cat}{CAT-probing}

\title{CAT-probing: A Metric-based Approach to Interpret How Pre-trained Models for Programming Language Attend Code Structure}

\author{Nuo Chen\thanks{~~Equal contribution, authors are listed alphabetically.}~, Qiushi Sun\footnotemark[1]~, Renyu Zhu\footnotemark[1]~, Xiang Li\thanks{~~Corresponding author.}~, Xuesong Lu, and Ming Gao \\
        School of Data Science and Engineering, East China Normal University, Shanghai, China \\
        \texttt{\{nuochen,qiushisun,renyuzhu\}@stu.ecnu.edu.cn}, \\ \texttt{\{xiangli,xslu,mgao\}@dase.ecnu.edu.cn}}

\begin{document}
\maketitle
\begin{abstract}
Code pre-trained models (CodePTMs) have recently demonstrated significant success in code intelligence. 
To interpret these models, 
some probing methods 
have been applied. 
However, 
these methods fail to consider the inherent characteristics of codes.
In this paper,
to address the problem,
we propose a novel probing method CAT-probing
to quantitatively interpret how CodePTMs attend code structure. 
We first 
denoise the input code sequences 
based on the token types pre-defined by the compilers
to filter those tokens
whose attention scores are too small.
After that,
we define a new metric CAT-score
to 
measure the commonality between 
the token-level attention scores generated in CodePTMs 
and
the pair-wise distances between corresponding AST nodes.
The higher the CAT-score, the stronger the ability of CodePTMs to capture code structure.
We conduct extensive experiments 
to integrate \cat\ with
representative CodePTMs 
for different programming languages.
Experimental results show 
the effectiveness of
\cat\ 
in CodePTM interpretation.
Our codes and data are publicly available at \url{https://github.com/nchen909/CodeAttention}. 
\end{abstract}

\section{Introduction}

\begin{figure*}[ht!]
    \centering
    \subfigure[A Python code snippet with its U-AST]{ 
        \includegraphics[width=6.9cm]{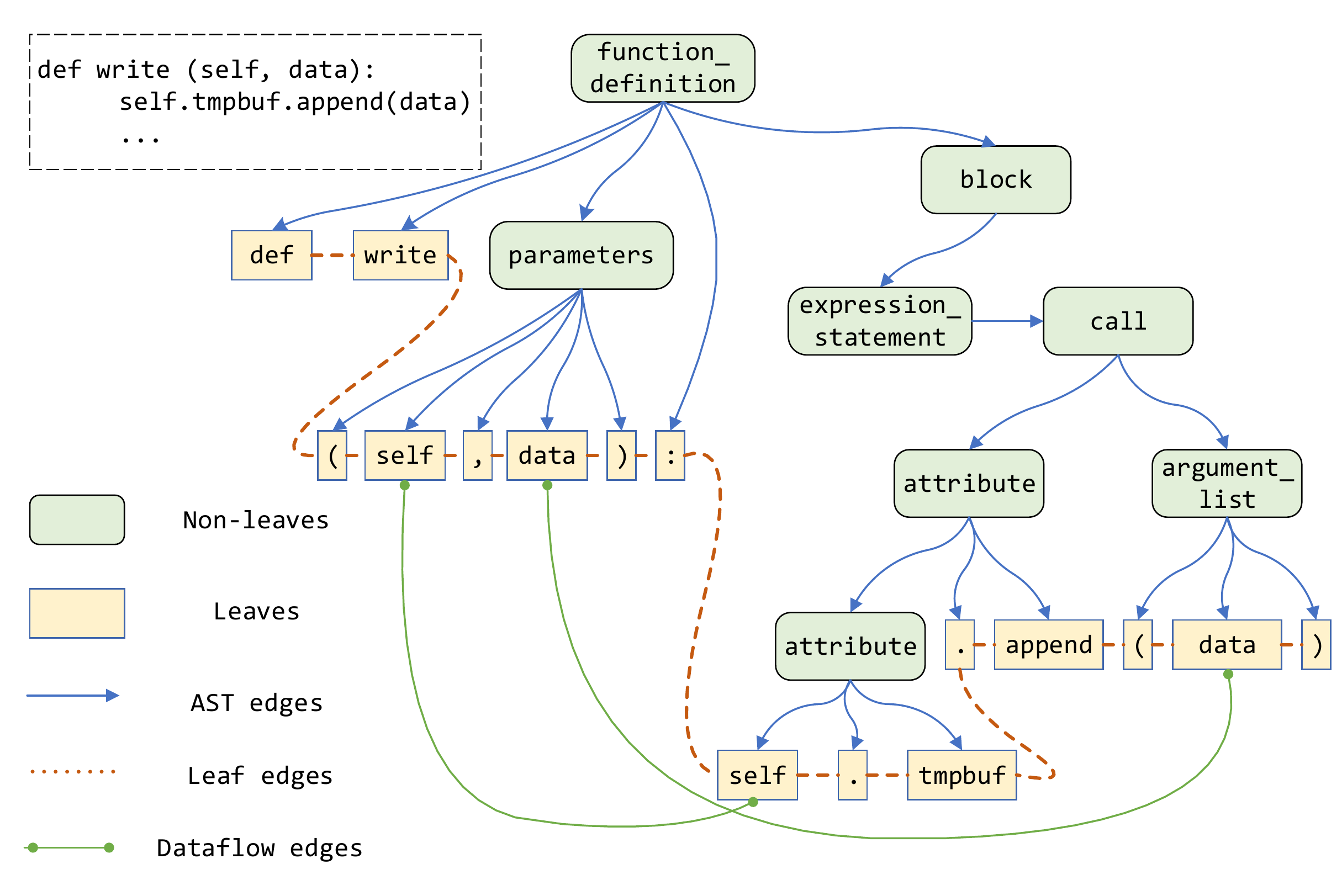}
        \label{pic1:ast-case}
    }
    \subfigure[The attention matrix (filtered)]{
	\includegraphics[width=4.12cm]{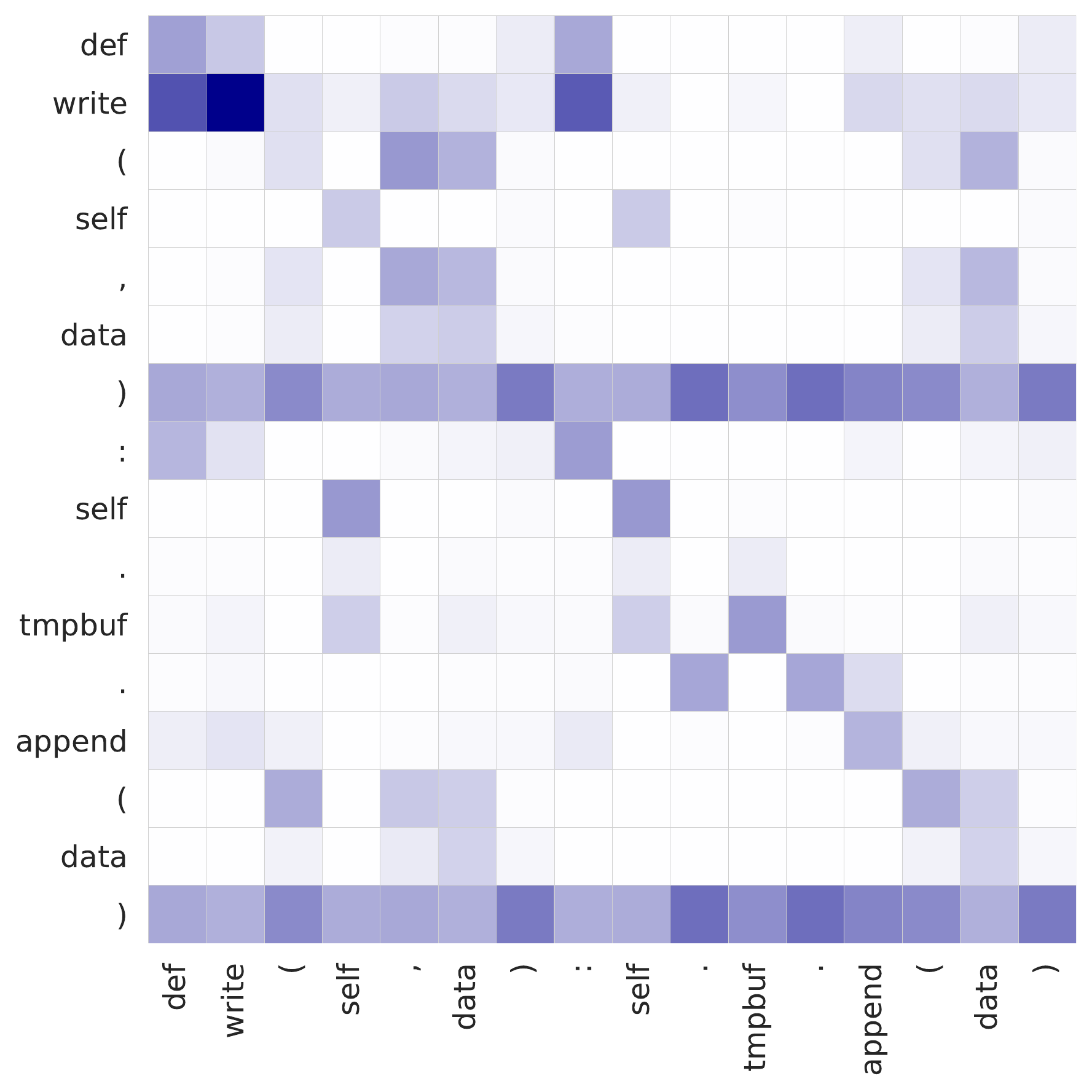}
	\label{pic1:py-att}
    }
    \subfigure[The distance matrix (filtered)]{
	\includegraphics[width=4.12cm]{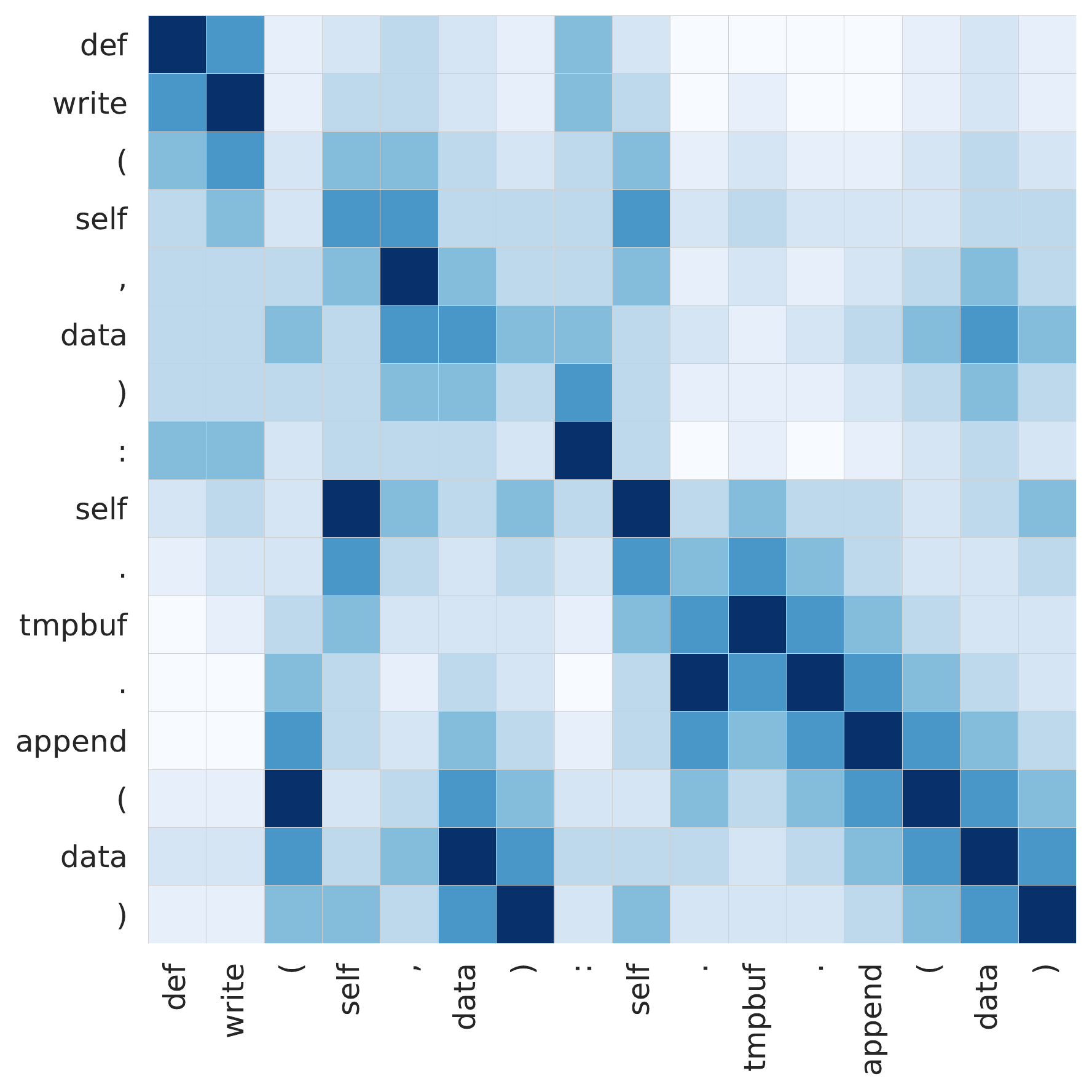}
	\label{pic1:py-dist}
	
    }
    \caption{Visualization 
    on the U-AST structure, 
    the attention matrix generated in the last layer of CodeBERT~\citep{feng-etal-2020-codebert} and the distance matrix. (a) A Python code snippet with its corresponding U-AST. 
    (b) Heatmaps of the averaged attention weights 
    after attention matrix filtering. (c) Heatmaps of the pair-wise token distance in U-AST.
    In the heatmaps,
    the darker the color,
    the more salient the attention score, or the closer the nodes.
    In this toy example, 
    only the token ``.'' between ``tmpbuf'' and ``append'' is filtered. More visualization examples of filtering are given in Appendix~\ref{sec:case-study}. 
    }
    \label{toktype}
\end{figure*}

In the era of ``Big Code''~\citep{allamanis2018survey},
the programming platforms, 
such as \textit{GitHub} and \textit{Stack Overflow},
have generated 
massive open-source code data.
With the assumption of ``Software Naturalness'' ~\cite{HindleBGS16}, 
pre-trained models~\cite{vaswani2017attention, devlin-etal-2019-bert, liu2019roberta} have been applied in the domain of code intelligence.

Existing code pre-trained models (CodePTMs) can be mainly divided into two categories:
\textit{structure-free} methods~\cite{feng-etal-2020-codebert, Svyatkovskiy2020IntelliCodeCC} 
and \textit{structure-based} methods~\citep{wang-etal-2021-codet5, niu2022spt}. 
The former
only utilizes the information from raw code texts,
while 
the latter employs 
code structures,
such as data flow~\citep{guo2021graphcodebert} and flattened AST\footnote{Abstract syntax tree.}~\citep{guo-etal-2022-unixcoder}, 
to enhance the performance of pre-trained models. For more details, readers can refer to
\citet{abs-2205-11739}.
Recently,
there exist works that 
use
probing techniques~\cite{clark-etal-2019-bert, vig-belinkov-2019-analyzing, zhang-etal-2021-sociolectal} 
to investigate what CodePTMs learn.
For example,
\citet{karmakar2021pre} first probe into CodePTMs and 
construct four probing tasks to explain them.
\citet{troshin2022probing} also
define 
a series of 
novel diagnosing probing tasks about code syntactic structure. 
Further,
\citet{ICSE22-Capture} conduct qualitative structural 
analyses 
to evaluate how CodePTMs interpret code structure. 
Despite the success,
all these methods 
lack quantitative characterization on
the degree of 
how well
CodePTMs learn from 
code structure.
Therefore, 
a research question arises: 
\textit{Can we develop a new probing way 
to evaluate how CodePTMs attend code structure quantitatively?}

In this paper, 
we propose a metric-based probing method,
namely, CAT-probing,
to quantitatively evaluate how \underline{C}odePTMs \underline{A}ttention scores relate to distances between AS\underline{T} nodes.
First,
to denoise the input code sequence in the original attention scores matrix, 
we classify the rows/cols by 
token types that are pre-defined by compilers,
and then 
retain tokens whose types
have the highest proportion scores 
to derive a filtered 
attention matrix (see Figure~\ref{pic1:py-att}). 
Meanwhile, 
inspired by the works~\citep{Detecting-Code-Clones, zhu-etal-2022-neural}, 
we add edges 
to improve the connectivity of AST and 
calculate the distances between 
nodes corresponding to the selected tokens,
which generates 
a distance matrix as shown in Figure~\ref{pic1:py-dist}. 
After that, 
we define {CAT-score}
to measure 
the matching degree between the filtered attention matrix and the distance matrix.
Specifically,
the point-wise elements of the two matrices are 
\emph{matched} if both the two conditions are satisfied: 
1) the attention score is larger than a threshold;
2) the distance value is smaller than a threshold.
If only one condition is reached, 
the elements are \emph{unmatched}.
We calculate the CAT-score 
by the ratio of the number of {matched} elements to the summation of {matched} and {unmatched} elements. 
Finally, the CAT-score is used to interpret how CodePTMs attend code structure, where a higher score indicates that the model has learned more structural information.

Our main contributions can be summarized as follows:

\begin{itemize}
    \item We 
    propose a novel metric-based probing method CAT-probing 
    to quantitatively interpret how CodePTMs attend code structure.
    
\item We apply CAT-probing to several representative CodePTMs and perform extensive experiments to demonstrate the effectiveness of our method (See Section~\ref{sec:cat-prob-effect}).

\item  
    We draw two fascinating observations from the empirical evaluation: 
    1) The token types that PTMs focus on vary with programming languages and are quite different from the general perceptions of human programmers (See Section~\ref{sec:fre-tok-typ}).
    2) The ability of CodePTMs to capture code structure dramatically differs with layers (See Section~\ref{sec:layerwise-analysis}).
    
\end{itemize}

\section{Code Background}

\subsection{Code Basics}

Each code can be represented in two modals: 
the source code and the code structure (AST), 
as shown in Figure~\ref{pic1:ast-case}. 
In this paper, 
we use Tree-sitter~\footnote{\href{https://github.com/tree-sitter}{github.com/tree-sitter}} to generate ASTs,
where
each token in the raw code 
is tagged with 
a unique type, 
such as ``identifier'', ``return''
and ``=''. 
Further,
following these works~\cite{Detecting-Code-Clones, zhu-etal-2022-neural}, 
we 
connect adjacent leaf nodes 
by adding data flow edges,
which increases the connectivity of AST. 
The upgraded AST is named as U-AST.

\subsection{Code Matrices}
\label{sec:code_mat}
There are two types of code matrices: 
the attention matrix and the distance matrix.
Specifically,
the attention matrix 
denotes the attention score generated by the Transformer-based CodePTMs, 
while the distance matrix captures 
the distance between nodes in U-AST.
We 
transform the original subtoken-level attention matrix into the token-level attention matrix by averaging the attention scores of subtokens in a token. 
For the distance matrix,
we use the shortest-path length to compute 
the distance between the leaf nodes of U-AST. 
Our attention matrix and distance matrix are shown in Figure~\ref{pic1:py-att} and Figure~\ref{pic1:py-dist}, respectively.

\section{CAT-probing}

\subsection{Code Matrices Filtering}
\label{sel-mat-gen}
As pointed out in~\cite{zhou2021informer},   
the attention scores in the attention matrix 
follow a long tail distribution,
which means that 
the majority of attention scores are very small.
To address the problem,
we propose a simple but effective algorithm based on code token types to remove the small values in the attention matrix. 
For space limitation, we summarize the pseudocodes of the algorithm in Appendix Alg.\ref{alg:fre_typ_sel}.
We only keep the rows/cols corresponding to frequent token types in the original attention matrix and distance matrix to generate the selected attention matrix and distance matrix.

\subsection{CAT-score Calculation}
After the two code matrices are filtered,
we define a metric called CAT-score, 
to measure the commonality between 
the filtered attention matrix $\mathbf{A}$ and 
the distance matrix $\mathbf{D}$. 
Formally,
the CAT-score is formulated as: 
\begin{equation}
\label{cat-score}
\text{CAT-score} = \frac{\sum_{C}\sum_{i=1}^{n}\sum_{j=1}^{n}\mathbbm{1}_{\mathbf{A}_{ij} >\theta_A \text{ and } \mathbf{D}_{ij} < \theta_D}}{\sum_{C}\sum_{i=1}^{n}\sum_{j=1}^{n}\mathbbm{1}_{\mathbf{A}_{ij} >\theta_A \text{ or } \mathbf{D}_{ij} < \theta_D}},
\end{equation}
where $C$ is the number of code samples, 
$n$ is the length of $\mathbf{A}$ or $\mathbf{D}$, $\mathbbm{1}$ is the indicator function, $\theta_A$ and $\theta_D$ denotes the thresholds to filter matrix $\mathbf{A}$ and $\mathbf{D}$, respectively. Specifically, we calculate the CAT-score of the last layer in CodePTMs. 
The larger the CAT-score, the stronger the ability of CodePTMs to attend code structure.

\section{Evaluation}

\subsection{Experimental Setup}

\paragraph{Task} 
We evaluate the efficacy of CAT-probing on code summarization, which is one of the most challenging downstream tasks for code representation.
This task aims to generate a natural language (NL) comment for a given code snippet, 
using smoothed BLEU-4 scores~\citep{lin-och-2004-orange} as the metric. 

\paragraph{Datasets}
We use the code summarization dataset from CodeXGLUE~\citep{CodeXGLUE2102-04664} to evaluate the effectiveness of our methods on four programming languages (short as PLs),  which are JavaScript, Go, Python and Java. For each programming language, we randomly select $C = 3,000$ examples from the training set for probing. 

\paragraph{Pre-trained models}
We select four models, including one PTM, namely RoBERTa~\citep{liu2019roberta}, and three RoBERTa-based CodePTMs, which are CodeBERT~\citep{feng-etal-2020-codebert}, 
GraphCodeBERT~\citep{guo2021graphcodebert}, and UniXcoder~\citep{guo-etal-2022-unixcoder}. 
All these PTMs are composed of $12$ layers of Transformer with $12$ attention heads. 
 We conduct layer-wise probing on these models, where the layer attention score is defined as the average of $12$ heads' attention scores in each layer.
The comparison of these models is introduced in Appendix~\ref{appd:cptms-overview}. And the details of the experimental implementation are given in Appendix~\ref{cat-probing-hyperparam}.


In the experiments,
we aim to answer the three research questions in the following:

\begin{itemize}
    \item \textbf{RQ1(Frequent Token Types)}: What kind of language-specific frequent token types do these CodePTMs pay attention to?
    \item \textbf{RQ2(CAT-probing Effectiveness)}: Is CAT-probing an effective method to evaluate how CodePTMs attend code structure?
    \item \textbf{RQ3(Layer-wise CAT-score)}: How does the CAT-score change with layers?
\end{itemize}

\begin{figure}[htbp]
    \centering
    \subfigure[Go]{ 
        \includegraphics[width=3.4cm]{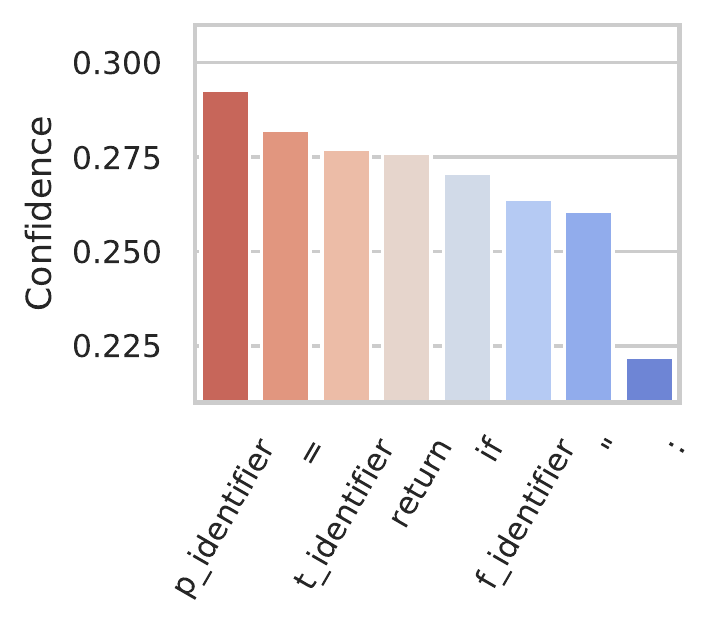}
    }
    \subfigure[Java]{
	\includegraphics[width=3.4cm]{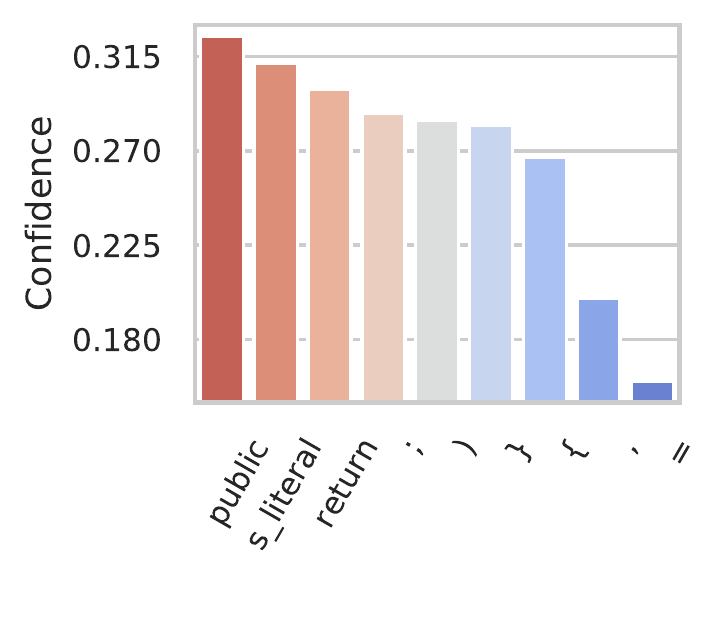}
    }
    \quad   
    \subfigure[JavaScript]{
    	\includegraphics[width=3.4cm]{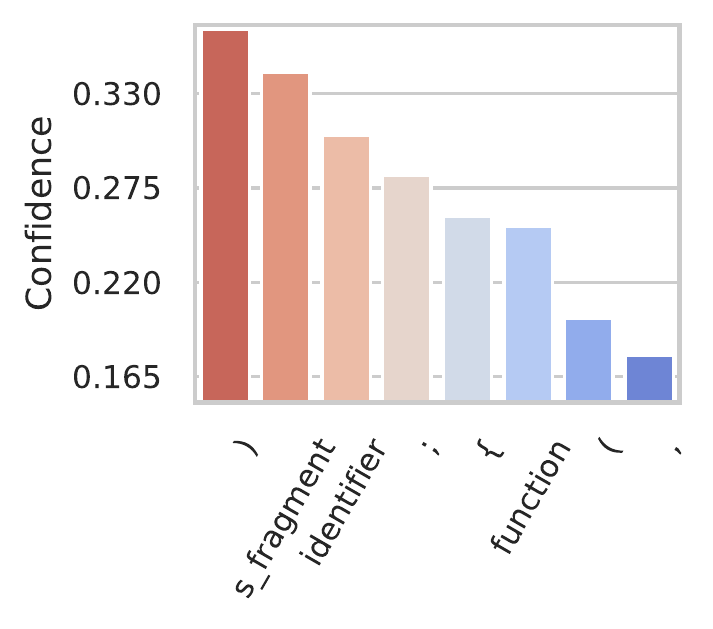}
    }
    \subfigure[Python]{
	\includegraphics[width=3.4cm]{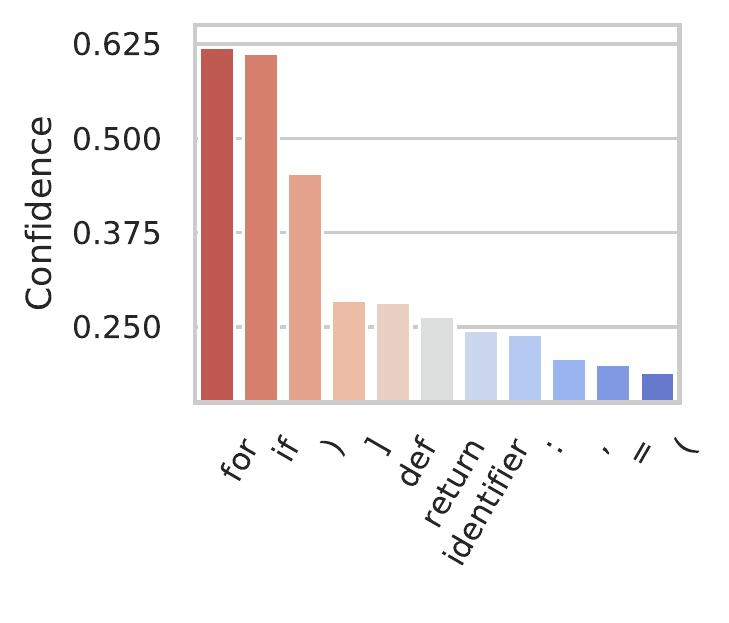}
    }
    \caption{Visualization of the frequent token types on four programming languages.}

    \label{toktype-barplot}
\end{figure}

\subsection{Frequent Token Types}
\label{sec:fre-tok-typ}

Figure~\ref{toktype-barplot}(a)-(d) demonstrates the language-specific frequent token types for four PLs, 
respectively. 
From this figure, 
we see that:
1) Each PL has its language-specific frequent token types and these types are quite different.
For example, 
the Top-3 frequent token types for Java are ``public'', ``s\_literal'' and ``return'', 
while Python are ``for'', ``if'', ``)''.
2) There is a significant gap between the frequent token types that CodePTMs focus on and the general perceptions of human programmers. For instance, CodePTMs assigned more attention to code tokens such as brackets.
3) Attention distribution on Python code snippets significantly differs from others. This is caused by Python having lesser token types than other PLs; thus, the models are more likely to concentrate on a few token types.

\begin{figure}
    \centering
    \subfigure[Go]{
        \includegraphics[width=3.605cm]{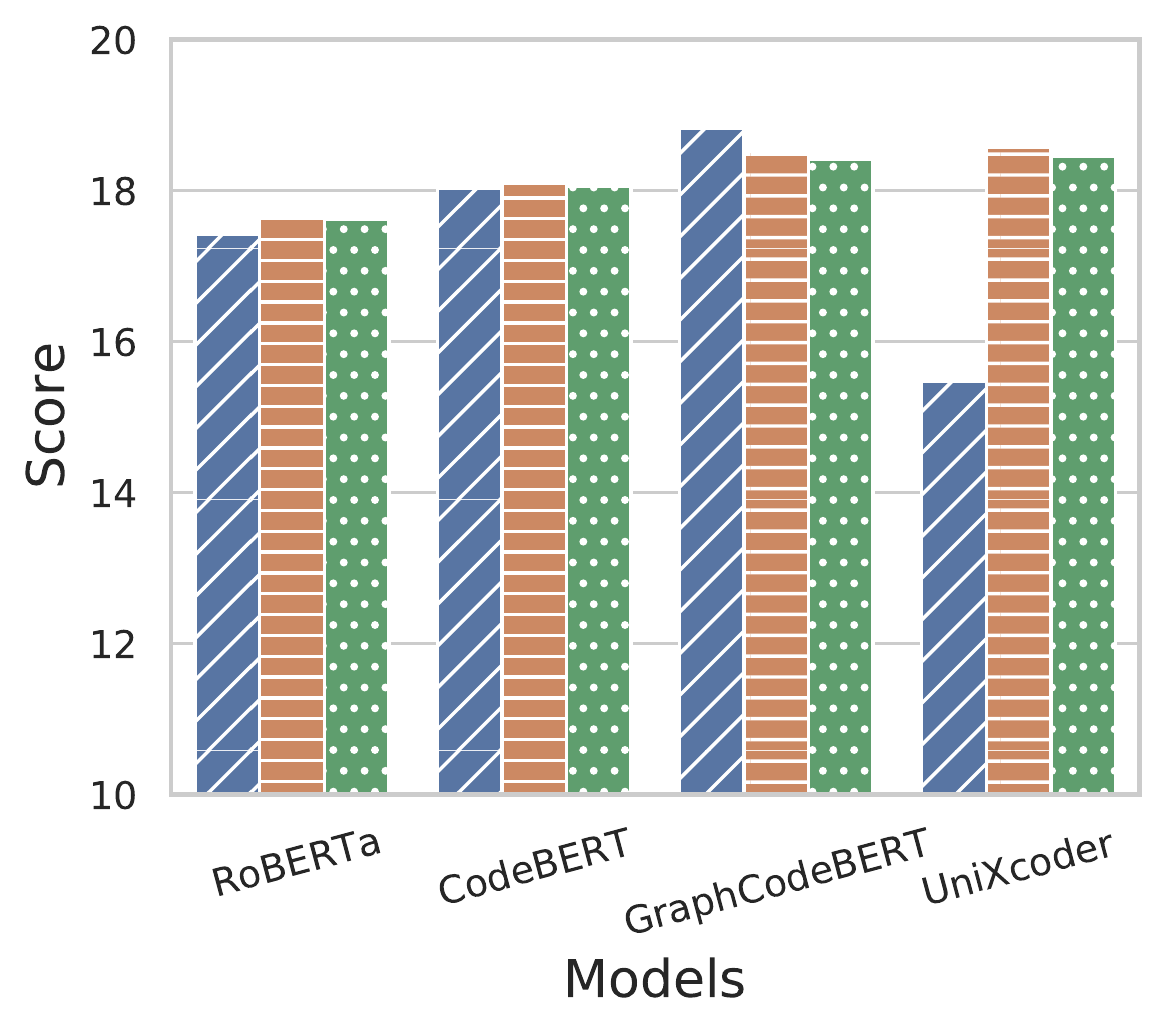}
        \label{sc-bleu:go}
    }
    \subfigure[Java]{
	\includegraphics[width=3.605cm]{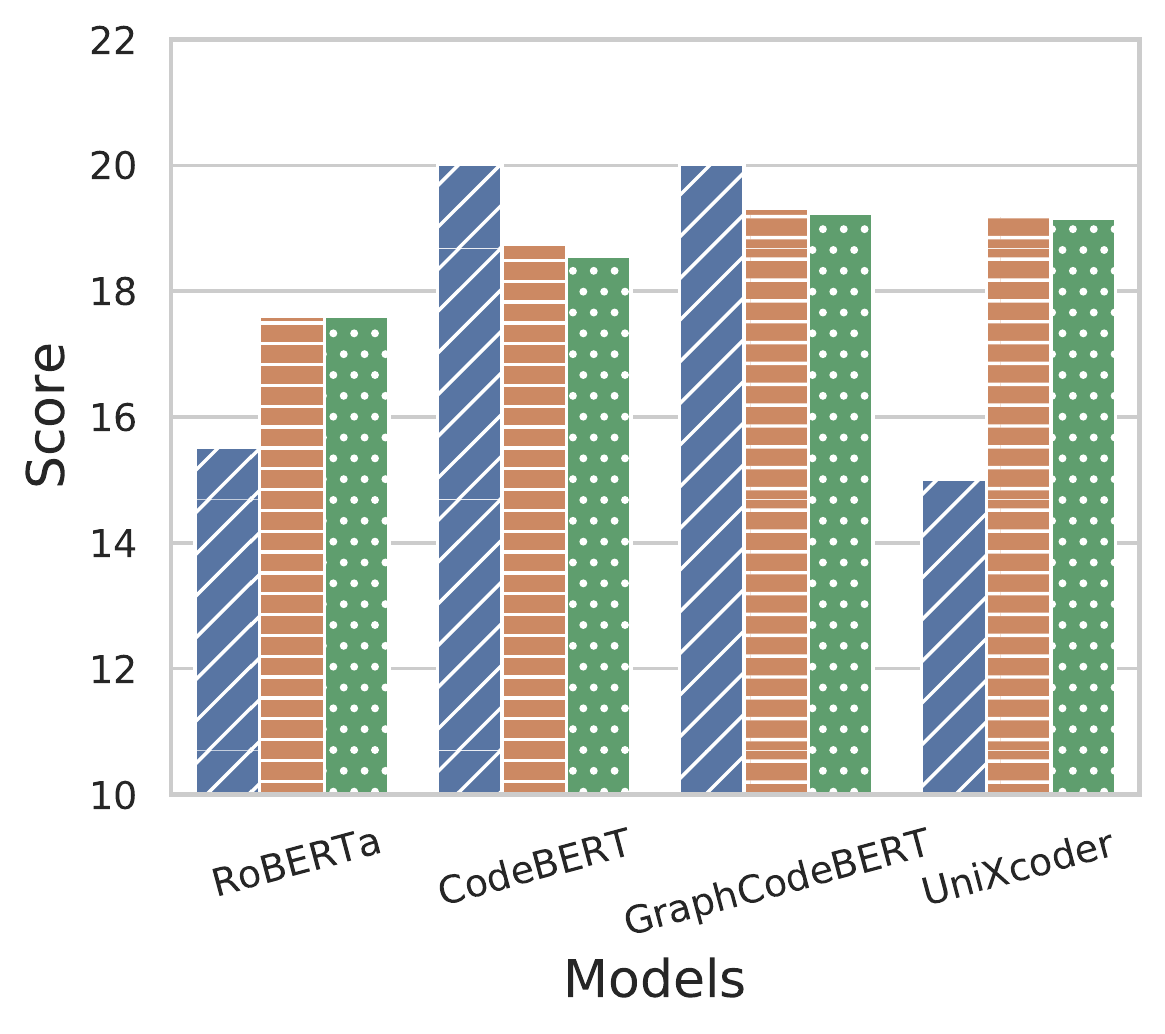}
	\label{sc-bleu:java}
    }
    \quad   
    \subfigure[JavaScript]{
    	\includegraphics[width=3.605cm]{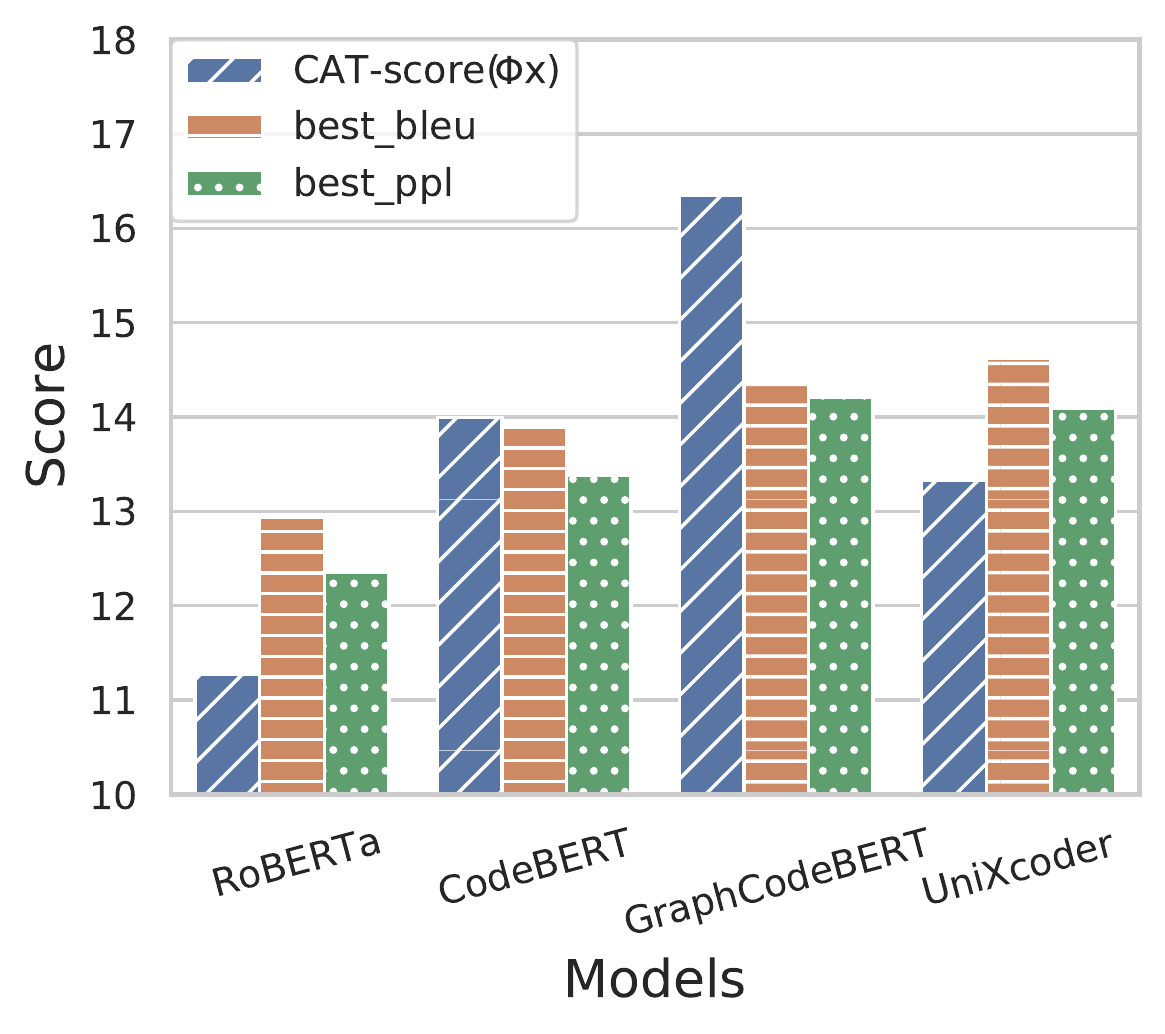}
    	\label{sc-bleu:js}
    }
    \subfigure[Python]{
	\includegraphics[width=3.605cm]{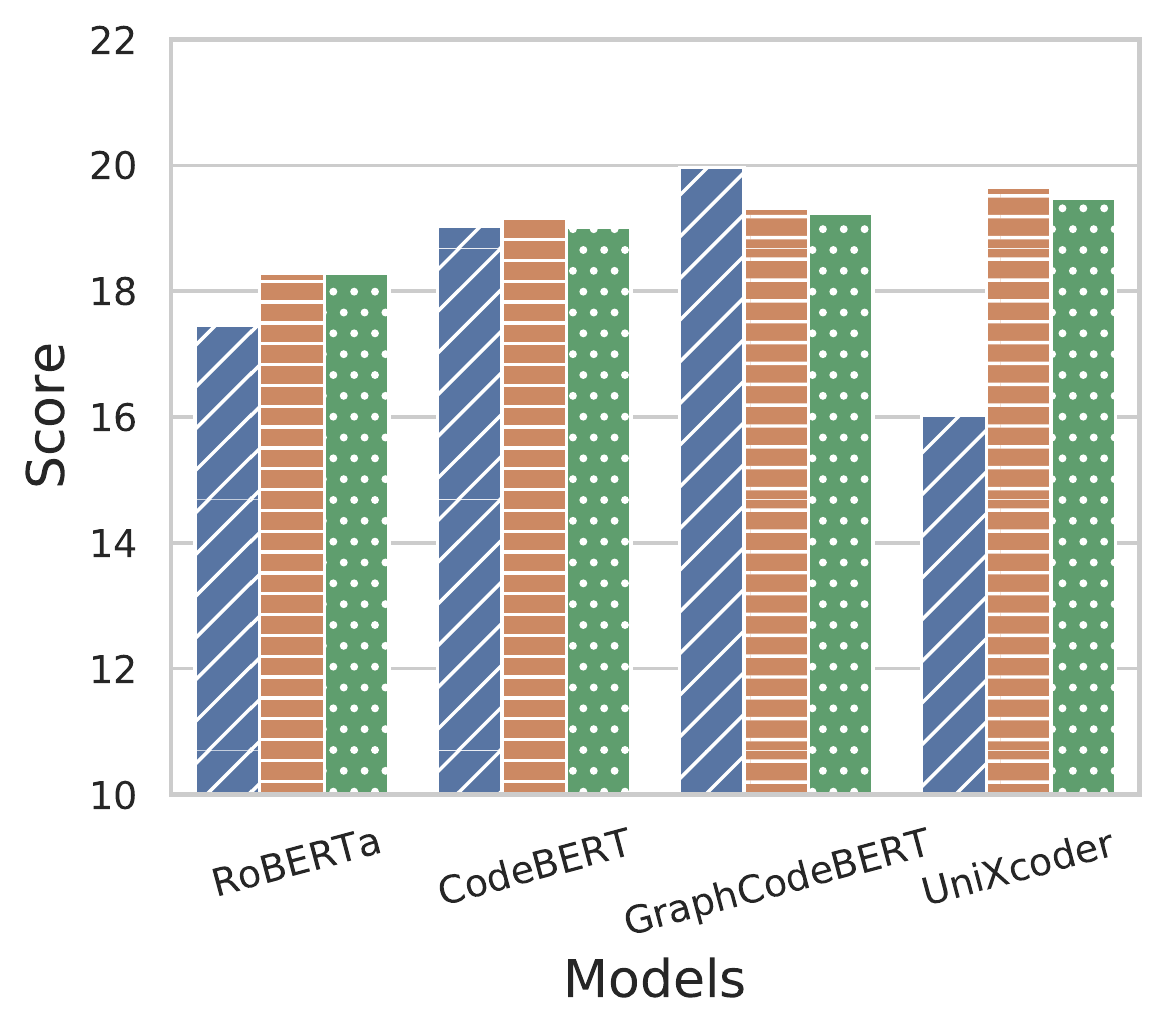}
	\label{sc-bleu:py}
    }
    \caption{Comparisons between the CAT-score and the performance on code summarization task. }
    \label{fig-score-bleu}
\end{figure}

\subsection{CAT-probing Effectiveness}
\label{sec:cat-prob-effect}

To verify the effectiveness of CAT-probing, we compare the CAT-scores with the models' performance on the test set (using both best-bleu and best-ppl checkpoints). The comparison among different PLs is demonstrated in Figure~\ref{fig-score-bleu}. We found strong concordance between the CAT-score and the performance of encoder-only models, including RoBERTa, CodeBERT, and GraphCodeBERT. This demonstrates the effectiveness of our approach in bridging CodePTMs and code structure. Also, this result (GraphCodeBERT > CodeBERT > RoBERTa) suggests that for PTMs, the more code features are considered in the input and pre-training tasks, the better structural information is learned.

In addition, we observe that UniXcoder has completely different outcomes from the other three CodePTMs. This phenomenon is caused by UniXcoder utilizing three modes in the pre-training stage (encoder-only, decoder-only, and encoder-decoder). 
This leads to a very different distribution of learned attention and thus different results in the CAT-score.

\subsection{Layer-wise CAT-score}
\label{sec:layerwise-analysis}
\begin{figure}
    \centering
    \subfigure[Go]{
        \includegraphics[width=3.605cm]{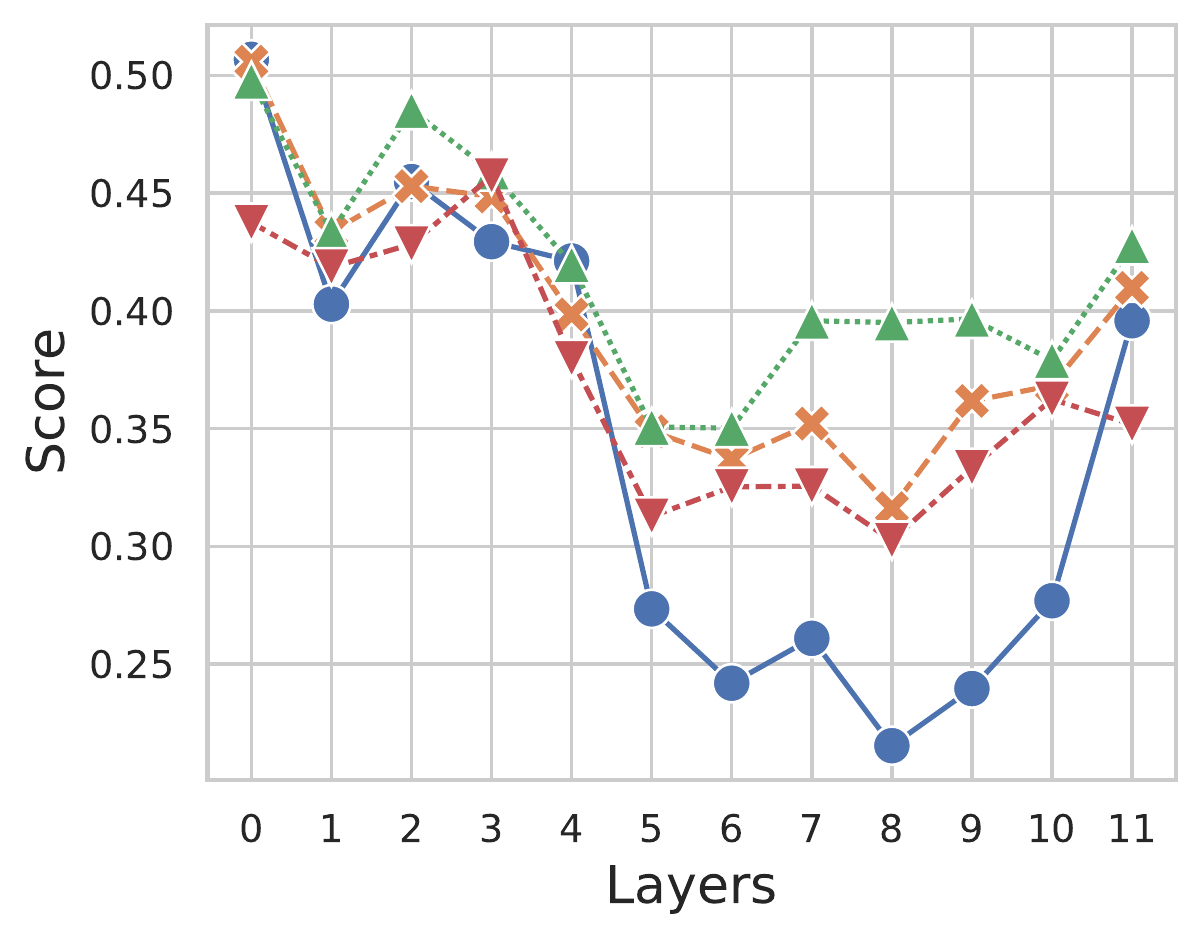}
        \label{layerwise:go}
    }
    \subfigure[Java]{
	\includegraphics[width=3.605cm]{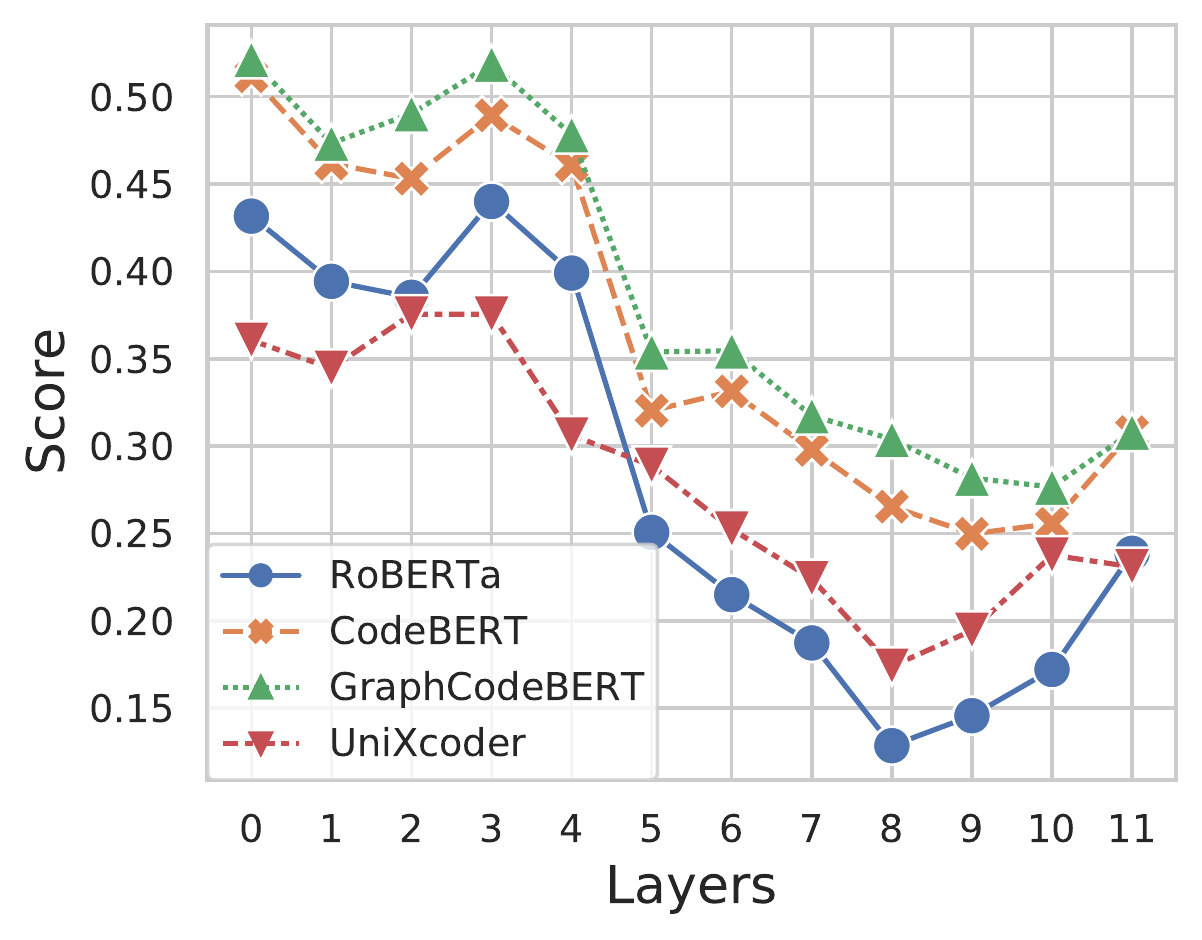}
	\label{layerwise:java}
    }
    \quad    
    \subfigure[JavaScript]{
    	\includegraphics[width=3.605cm]{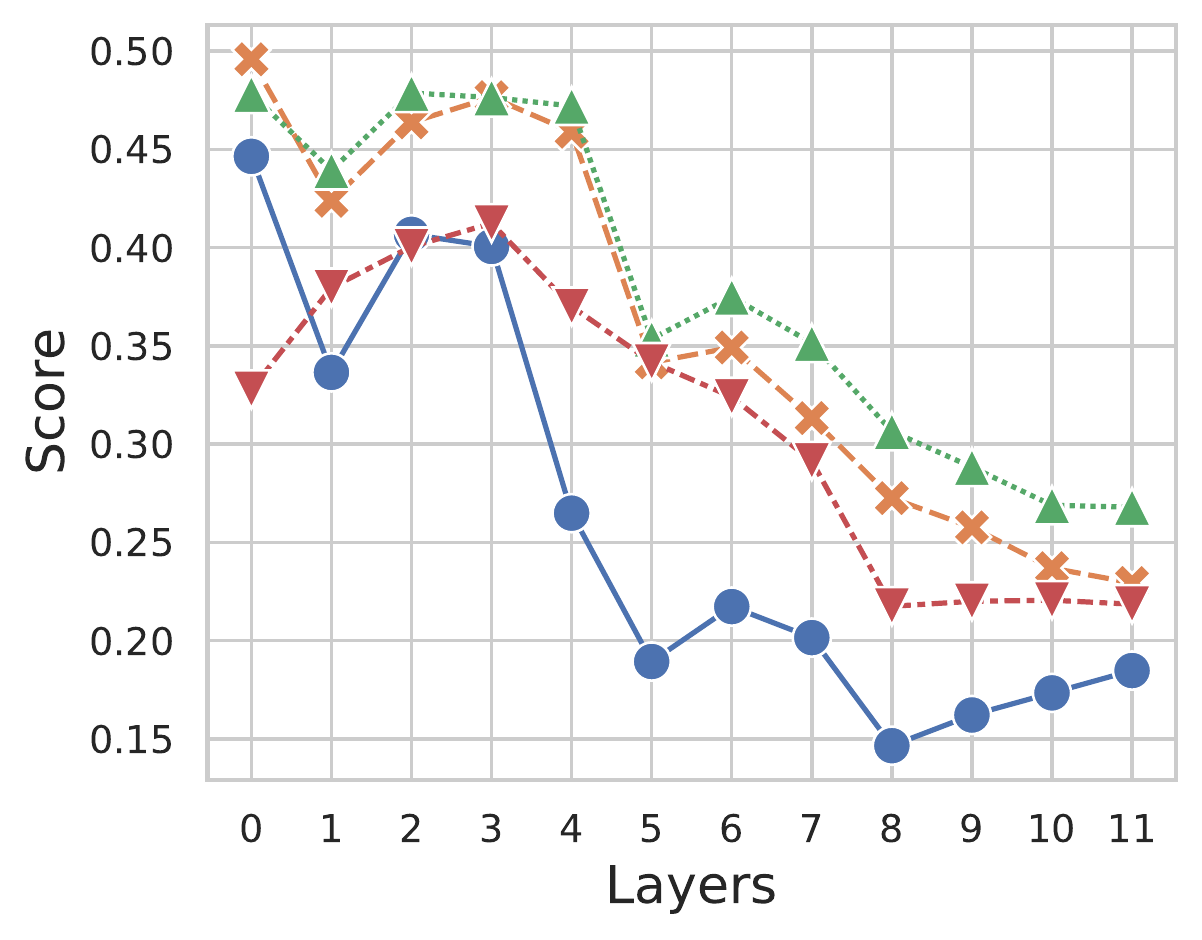}
    	\label{layerwise:js}
    }
    \subfigure[Python]{
	\includegraphics[width=3.605cm]{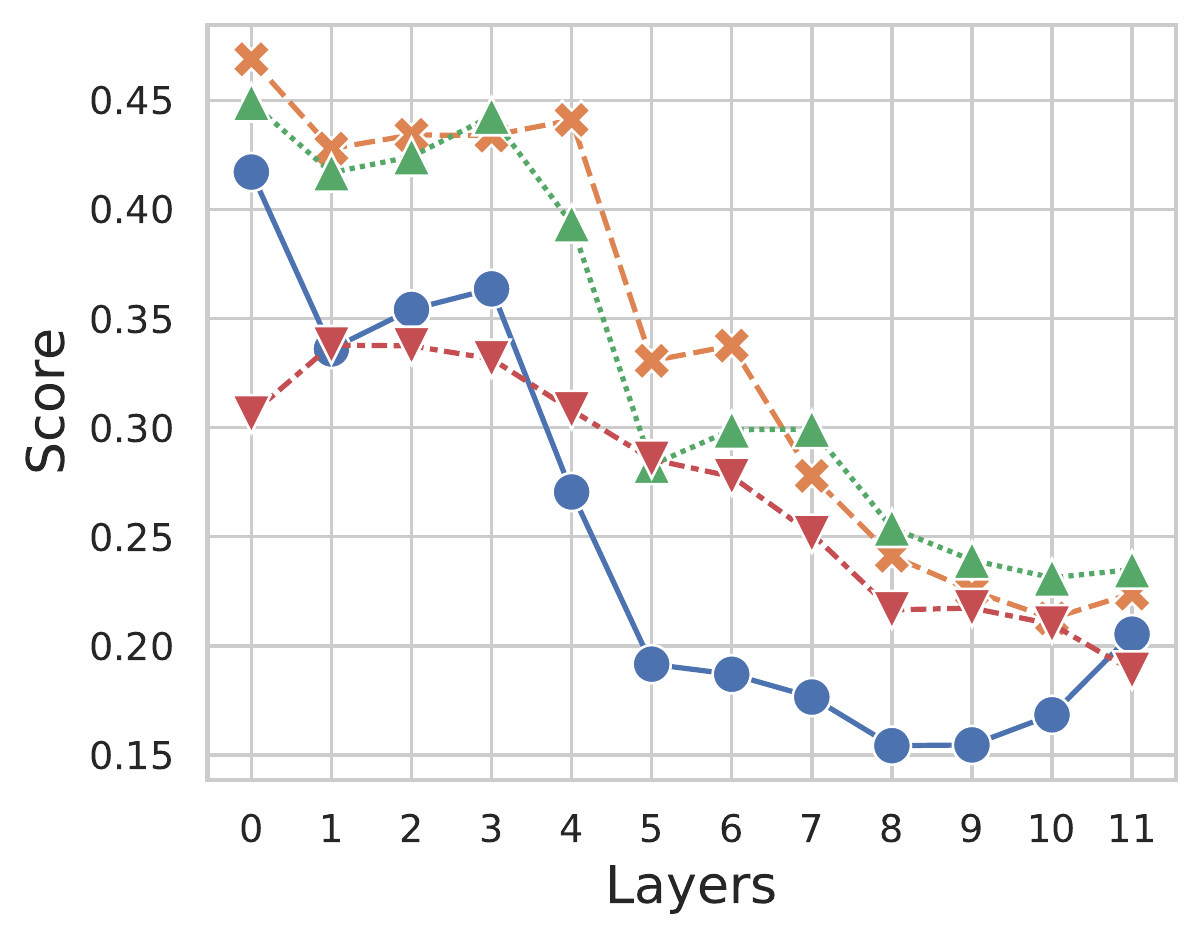}
	\label{layerwise:Py}
    }
    \caption{Layer-wise CAT-score results.}
    \label{fig:score-each-layer}
\end{figure}

We end this section with a study on layer-wise CAT-scores. 
Figure~\ref{fig:score-each-layer} gives the results of the CAT-score on all the layers of PTMs.
From these results, 
we observe that:
1) The CAT-score decreases in general when the number of layers increases on all the models and PLs.
This is because attention scores gradually focus on some special tokens, 
reducing the number of matching elements.
2) The relative magnitude relationship (GraphCodeBERT > CodeBERT > RoBERTa) between CAT-score is almost determined on all the layers and PLs,
which indicates the effectiveness of CAT-score to recognize the ability of CodePTMs in capturing code structure.
3) In the middle layers (4-8), all the results of CAT-score change drastically, which indicates the middle layers of CodePTMs may play an important role in transferring general structural knowledge into task-related structural knowledge. 
4) In the last layers (9-11), CAT-scores gradually converge, i.e.,  the models learn the task-specific structural knowledge, which explains why we use the score at the last layer in CAT-probing.

\section{Conclusion}

In this paper, 
we proposed a novel probing method named CAT-probing to explain how CodePTMs attend code structure. 
We first 
denoised the input code sequences 
based on the token types pre-defined by the compilers
to filter those tokens
whose attention scores are too small.
After that,
we defined a new metric CAT-score
to 
measure the commonality between 
the token-level attention scores generated in CodePTMs 
and
the pair-wise distances between corresponding AST nodes.
Experiments on multiple programming languages demonstrated the effectiveness of our method. 

\section{Limitations}
The major limitation of our work is that the adopted probing approaches mainly focus on encoder-only CodePTMs, which could be just one aspect of the inner workings of CodePTMs. In our future work, we will explore more models with encoder-decoder architecture, like CodeT5~\citep{wang-etal-2021-codet5} and PLBART~\citep{ahmad-etal-2021-unified}, and decoder-only networks like GPT-C~\citep{Svyatkovskiy2020IntelliCodeCC}.

\section*{Acknowledgement}
\label{sec:acknowledge}
This work has been supported by the National Natural Science Foundation of China under Grant No. U1911203, 
the National Natural Science Foundation of China under Grant No. 62277017,
Alibaba Group through the Alibaba Innovation Research Program, 
and the National Natural Science Foundation of China under Grant No. 61877018,
The Research Project of Shanghai Science and Technology Commission (20dz2260300) and The Fundamental Research Funds for the Central Universities.
And the authors would like to thank all the anonymous reviewers for their constructive and insightful comments on this paper.

\bibliographystyle{acl_natbib}
\bibliography{anthology, emnlp2022} 

\appendix

\section{Frequent Token Types Filtering Algorithm}
Algorithm~\ref{alg:fre_typ_sel} describes the procedure to generate frequent token types.

\begin{algorithm*}[t]
	\caption{Frequent Token Type Selection}
	\begin{algorithmic}[1]
		\Require \
		Language $lang$
		\Ensure \
		Frequent token type list $type\_list$
        \State rank = len(token types) * [0]
        \Comment{Initialize rank for each token type}
		\For{$t$ in token types}
		\For{$m$ in CodePTM models}
		\State confidence[$t$,$m$] = 0
		\For{$c$ in code cases}
		\State $att$ = get\_att($m$,$lang$,$c$)
		\Comment{Get attention matrix}
		\State $mask\_theta$ = is\_gt\_theta($att$)
		\Comment{Set $att$ position greater than $\theta_A$ to 1, otherwise 0}
		\State $mask\_type$ = is\_type\_t($att$)
		\Comment{Set $att$ position is type $t$ to 1, otherwise 0}
		\State $part$ = sum\_mat($mask\_theta \& mask\_type$)
		\Comment{Sum all elements of the matrix}
		\State $overall$ = sum\_mat($mask\_type$)
		\State confidence[$t$,$m$] $\leftarrow$  confidence[$t$,$m$] + $part$ / $overall$
		\Comment{Compute confidence}
		\EndFor
		\State confidence[$t$,$m$] $\leftarrow$ confidence[$t$,$m$] / len($c$)
		\Comment{Average confidence}
		\State rank[$t$] $\leftarrow$ rank[$t$] + get\_rank(confidence,$m$)
		\Comment{Rank confidence for $m$, and sum rank for $t$}
		\EndFor
		\EndFor
        \renewcommand{\algorithmicrequire}{ \textbf{Return:}}
	    \Require \ 
		token type list includes those $t$ with rank[$t$]<40
	\end{algorithmic}
		\label{alg:fre_typ_sel}
\end{algorithm*}
\section{Comparison of CodePTMs}
\label{appd:cptms-overview}
Table~\ref{tab:cptms} gives the comparison of the PTMs used in our experiments from three perspectives: the inputs of the model, the pre-training task, and the training mode.


\section{Experimental Implementation}

\label{cat-probing-hyperparam}
We keep the same hyperparameter setting for all CodePTMs.
The detailed hyperparameters are given in Table~\ref{cat-probing-hyperparam-table}.

Our codes are implemented based on PyTorch.
All the experiments were conducted on a Linux server with two interconnected NVIDIA-V100 GPUs.

\begin{table}[H]
\centering
\begin{tabular}{cc}
\toprule
\textbf{Hyperparameter} & \textbf{value} \\
\midrule
Batch Size & 48 \\
Learning Rate & 5e-5 \\
Weight Decay &  0.0 \\
Epsilon &  1e-8 \\
Epochs & 15 \\
Max Source Length & 256 \\
$\theta_A$ & third quartile of values in $A$\\
$\theta_D$ & first quartile of values in $D$\\
\bottomrule
\end{tabular}
\caption{\label{cat-probing-hyperparam-table}
Hyperparameters for CAT-probing
}
\end{table}

\begin{center}

\begin{table*}[t]
\centering
\small
\begin{tabular}{lccc}
\toprule  
\textbf{Models} & \textbf{Inputs} &  \textbf{Pre-training Tasks} & \textbf{Training Mode}\\ \midrule
RoBERTa & Natural Language (NL) & Masked Language Modeling (MLM) & Encoder-only  \\
\midrule
CodeBERT & NL-PL Pairs & MLM+Replaced Token Detection (RTD) & Encoder-only \\
\midrule
GraphCodeBERT & NL-PL Pairs \& AST & MLM+Edge Prediction+Node Alignment & Encoder-only  \\
\midrule
\multirow{3}{*}{UniXcoder} & \multirow{3}{*}{NL-PL Pairs \& Flattened AST} &  MLM & Encoder \&\\
& & ULM (Unidirectional Language Modeling) & Decoder \&\\
& & Denoising Objective (DNS) & Encoder-decoder  \\ 
\bottomrule 
\end{tabular}
\caption{The comparison of different language models mentioned in this paper.}
\label{tab:cptms}
\end{table*}
\label{sec:appendix}
\end{center}

\newpage

\section{Case Study}
\label{sec:case-study}
In addition to the example visualized in Figure~\ref{toktype}, we have carried out three new examples to show the effectiveness of the filtering strategy in Section~\ref{sel-mat-gen}, The visualizations are shown in Table~\ref{tab:case-study}.

\begin{table*}[t]
\centering
\small
\begin{tabular}{lcc}
\toprule  
\multirow{2}{*}{\textbf{Source Code}} & \multirow{2}{*}{\textbf{Attention Heatmap}} & \textbf{Attention Heatmap with } \\ 
& & \textbf{Token Type Selection} \\
\midrule

\raisebox{0.75\height}{\includegraphics[width=4.5cm]{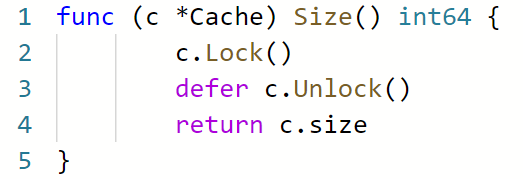}}
 & \includegraphics[width=4.5cm]{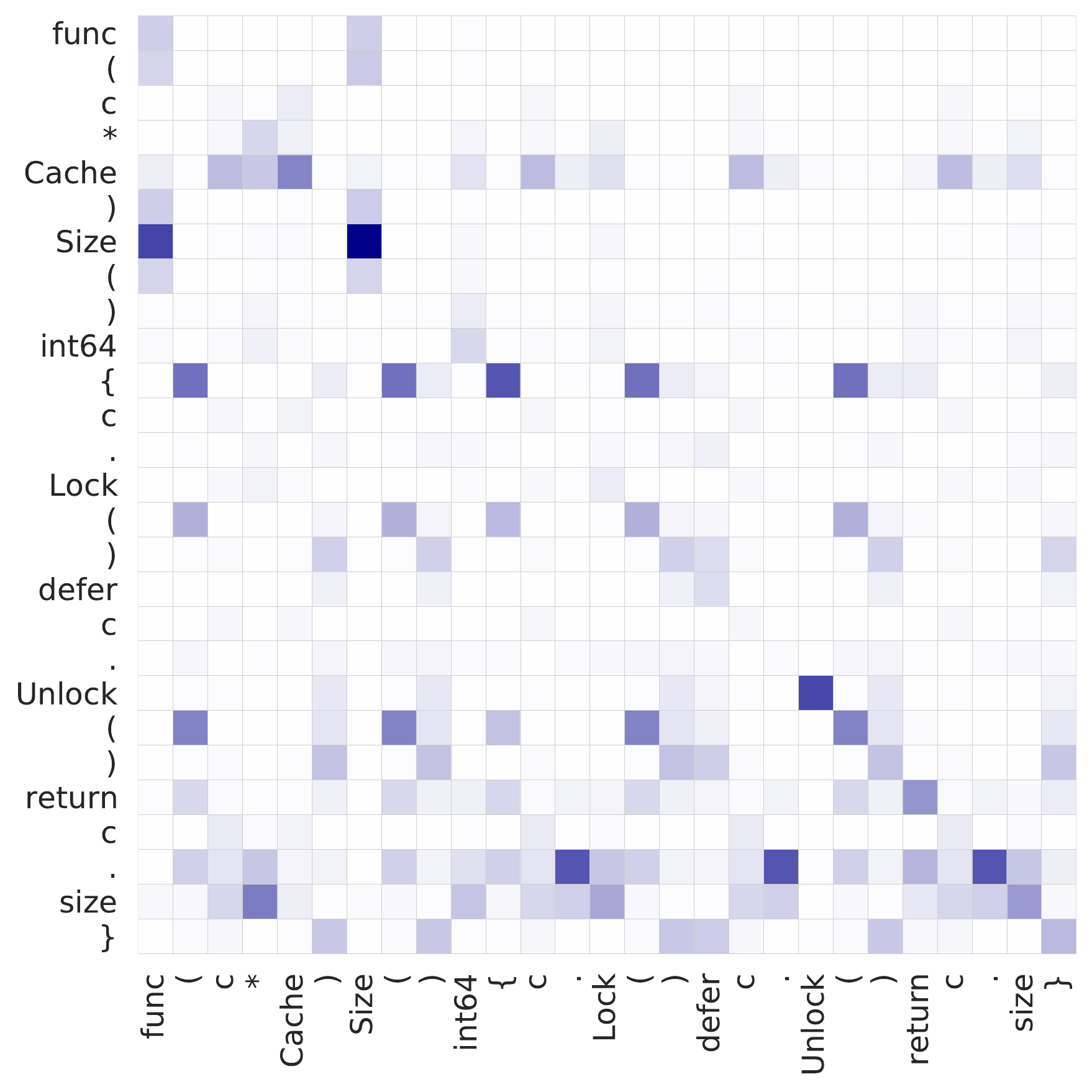} & \includegraphics[width=4cm]{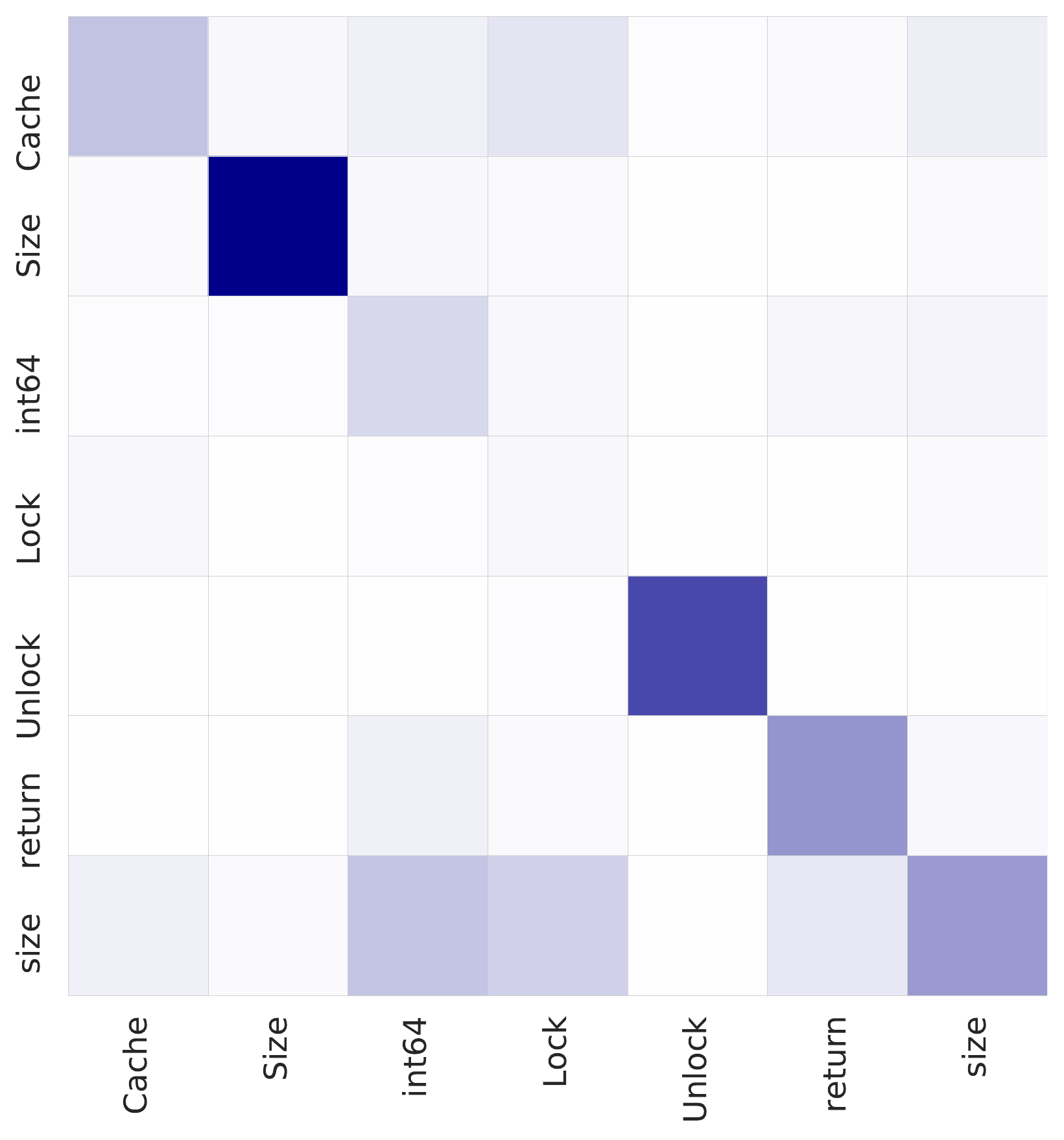} \\
\midrule
\raisebox{0.9\height}{\includegraphics[scale=0.25]{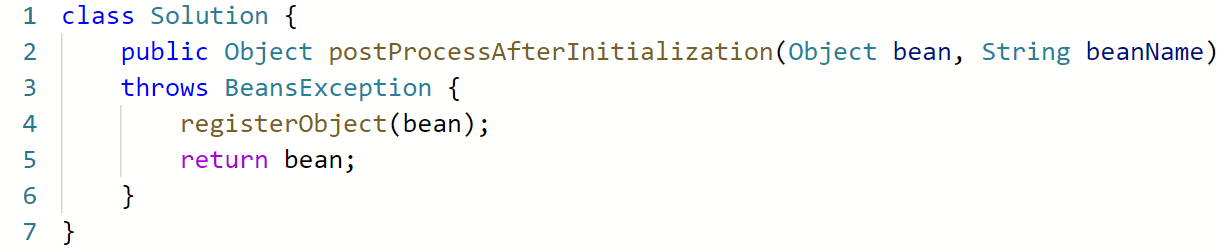}} & \includegraphics[width=4.5cm]{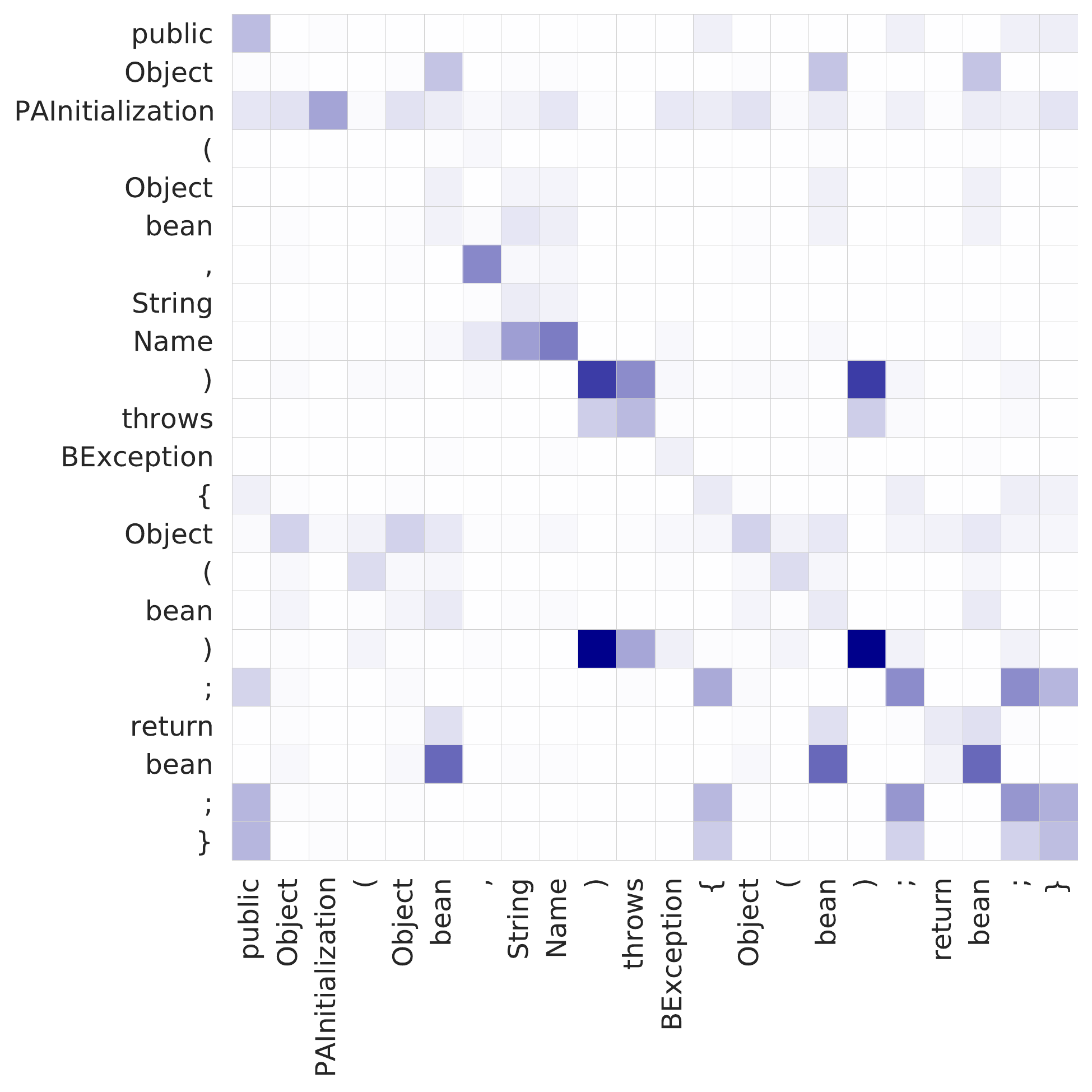} & \includegraphics[width=4cm]{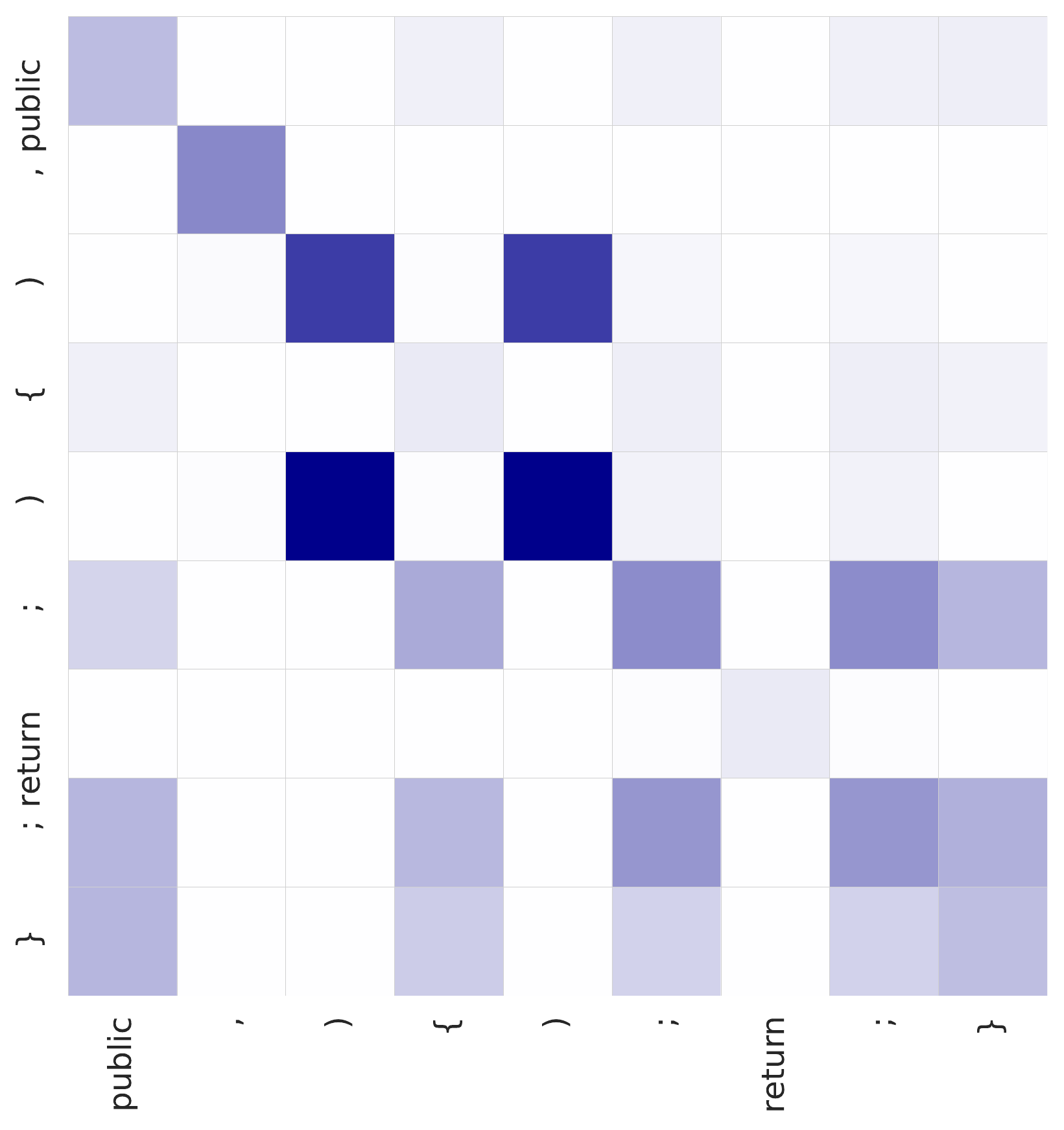}  \\
\midrule
\raisebox{0.95\height}{\includegraphics[scale=0.35]{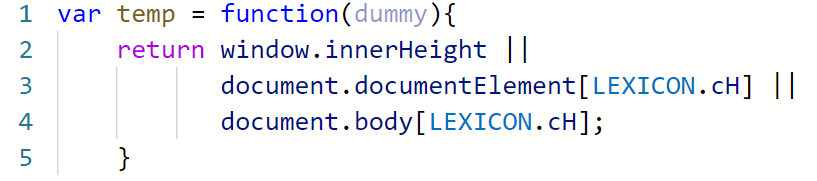}} & \includegraphics[width=4.5cm]{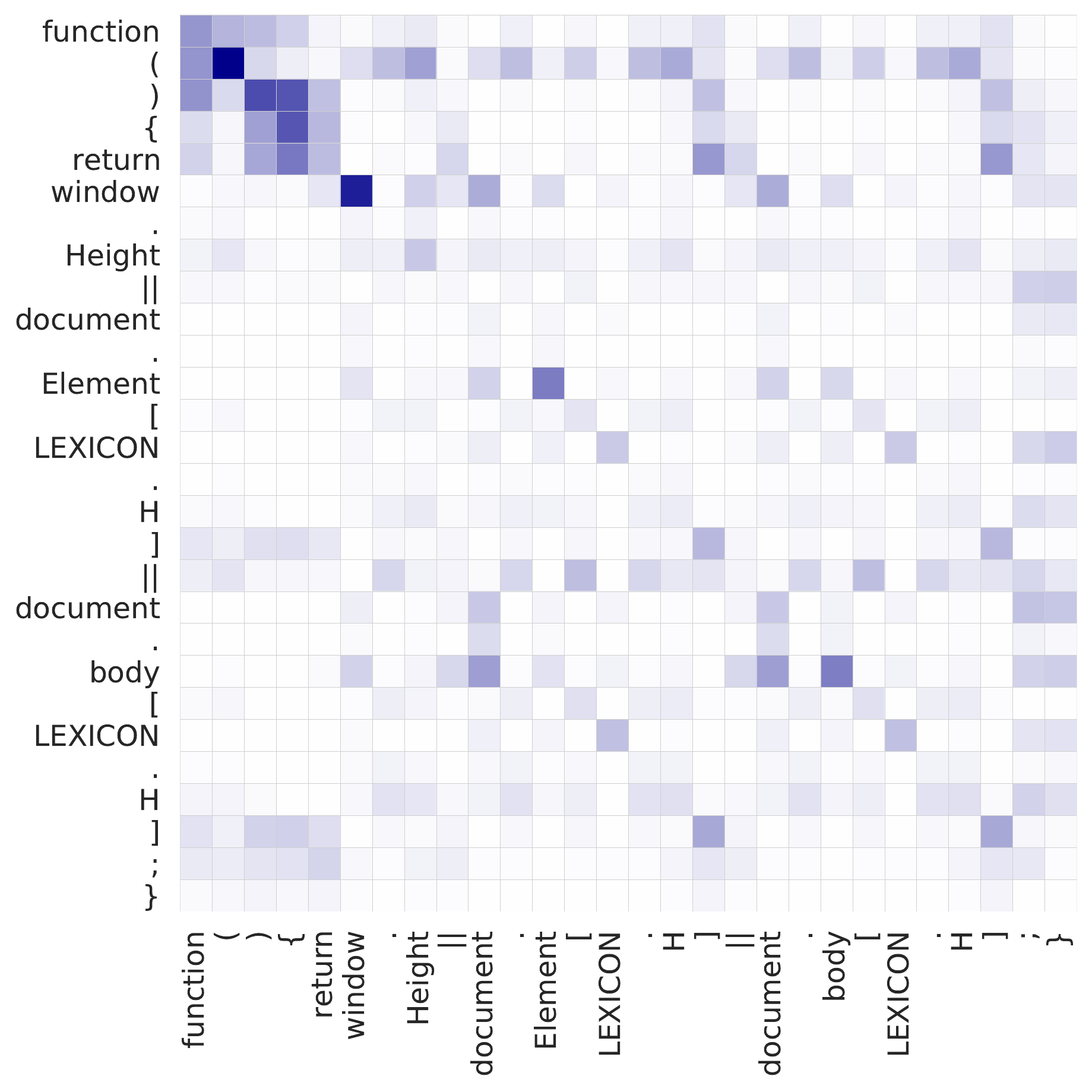}
& \includegraphics[width=4cm]{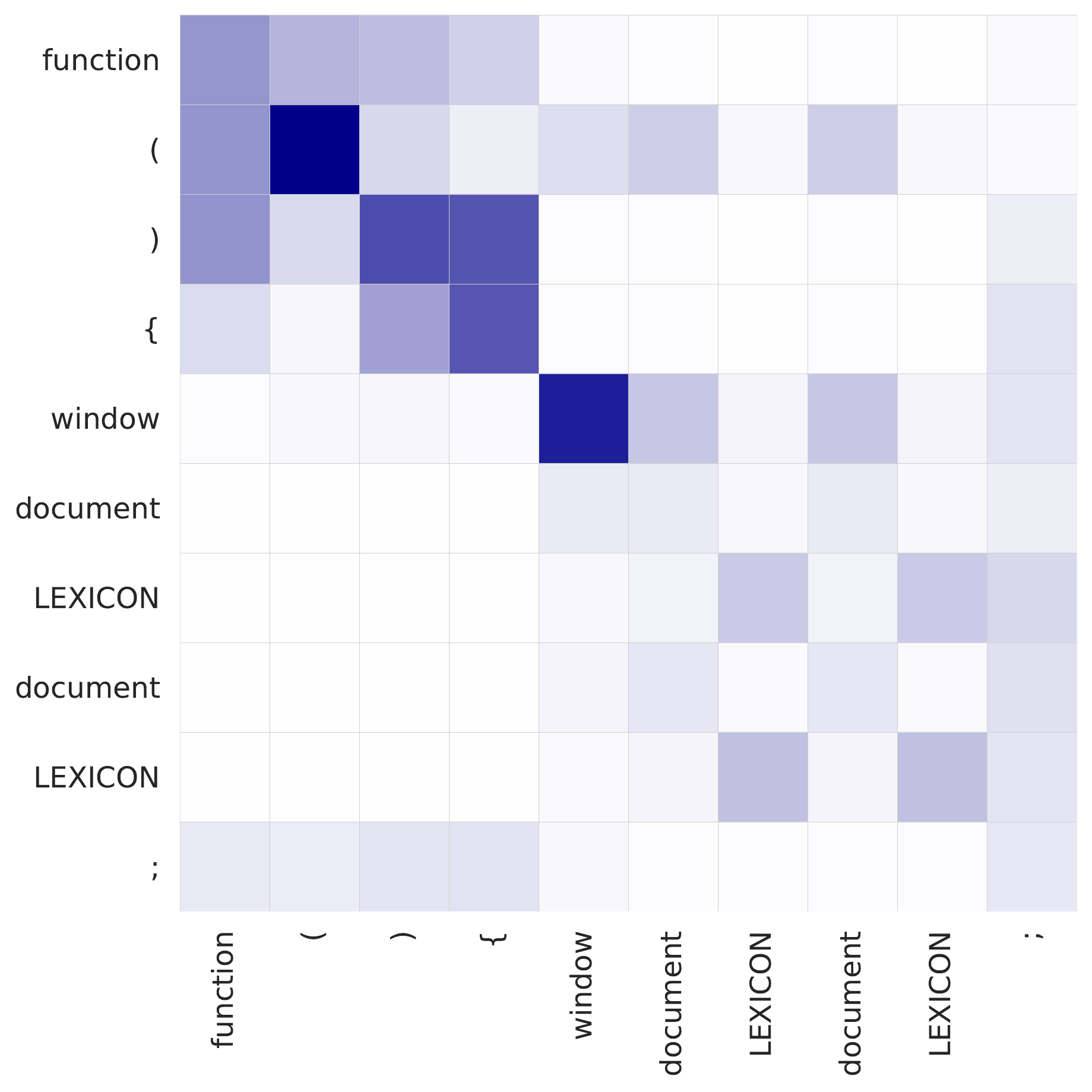}  \\
\bottomrule
\end{tabular}
\caption{Heatmaps of the averaged attention weights in the last layer before and after using token selection, including Go, Java, and JavaScript code snippets (from top to bottom).}
\label{tab:case-study}
\end{table*}

\end{document}